\documentclass{article_saj}
\usepackage{graphicx,saj,multicol,subeqnarray}
\usepackage{natbib}
\usepackage{float}
\usepackage{xcolor}
\usepackage{widetext}
\usepackage{url}
\usepackage{bm}
\usepackage{tikz} 
\usepackage{pifont} 
\usepackage{amsfonts}
\usepackage{amssymb}
\usepackage{amsmath,upgreek}
\usepackage{titlesec}
\titlelabel{\thetitle.\quad}
\definecolor{xlinkcolor}{cmyk}{1,0.6,0,0}
\usepackage[bookmarks=false,         
     pdfnewwindow=true,      
     colorlinks=true,    
     linkcolor=xlinkcolor,     
     citecolor=xlinkcolor,     
     filecolor=xlinkcolor,  
     urlcolor=xlinkcolor,      
final=true
]{hyperref}

\pdfoutput=1

\def\papertype{\ \hfill\ Editorial}

\def\udc{52}
\begin{document}
\parindent=.5cm
\baselineskip=3.8truemm
\columnsep=.5truecm
\newenvironment{lefteqnarray}{\arraycolsep=0pt\begin{eqnarray}}
{\end{eqnarray}\protect\aftergroup\ignorespaces}
\newenvironment{lefteqnarray*}{\arraycolsep=0pt\begin{eqnarray*}}
{\end{eqnarray*}\protect\aftergroup\ignorespaces}
\newenvironment{leftsubeqnarray}{\arraycolsep=0pt\begin{subeqnarray}}
{\end{subeqnarray}\protect\aftergroup\ignorespaces}
%


\markboth{\eightrm MODELING QUASAR VARIABILITY THROUGH SELF-ORGANIZING MAP-BASED NEURAL PROCESS} 
{\eightrm I. {\v C}VOROVI{\' C} - HAJDINJAK$^{1}$}

\begin{strip}

{\ }

\vskip-1cm

\publ

\type

{\ }


\title{MODELING QUASAR VARIABILITY THROUGH SELF-ORGANIZING MAP-BASED NEURAL PROCESS}


\authors{I. {\v C}vorovi{\' c} - Hajdinjak$^{1}$}

\vskip3mm


\address{$^1$Department of Astronomy, Faculty of Mathematics,
University of Belgrade\break Studentski trg 16, 11000 Belgrade,
Serbia}


\Email{iva.cvorovic@gmail.com}


\dates{May 18, 202x}{June 1, 202x}


\summary{Conditional Neural Process (QNPy) has shown to be a good tool for modeling quasar light curves. However, given the complex nature of the source and hence the data represented by light curves, processing could be time-consuming. In some cases, accuracy is not good enough for further analysis. In an attempt to upgrade QNPy, we examine the effect of the prepossessing quasar light curves via the Self-Organizing Map (SOM) algorithm on modeling a large number of quasar light curves. After applying SOM on SWIFT/BAT data and modeling curves from several clusters, results show the Conditional Neural Process performs better after SOM clustering. We conclude that SOM clustering of quasar light curves could be a beneficial prepossessing method for QNPy.}



\end{strip}

\tenrm


\section{INTRODUCTION}

\indent
Quasars are the luminous central regions of galaxies. \citep{lama15}. Their emission lines are observed at all wavelengths from radio to gamma rays. In all parts of electromagnetic spectra variability is strong, ranging from a few minutes to months \citep{zha24}. Optical lines are produced in clouds of gas, rapidly moving in the potential of the black hole - broad-line region \citep{urp95}. Analyzing optical variability in quasars has allowed us to understand the underlying physical processes  \citep{wag95, ulr03} and the structure of quasars \citep{ant93, rich06, bonf10, sha11}. Spectroscopic analyses have been instrumental in investigating the composition and kinematics of quasars. Complementary, quasar light curves provide an integrated temporal spectrum, capturing the aggregate spectral output as a function of time. These temporal profiles reflect the flux variations due to the dynamic astrophysical processes occurring in the accretion disk and surrounding environments of supermassive black holes. Therefore, analyzing these light curves is important for understanding the stochastic and transient events that characterize quasar variability which is essential for constraining the physical models that describe their emission mechanisms.

Quasar light curves are nonlinear, irregular, sparse, and could have gaps and signatures of extreme flares or noise. Such a structure could be challenging to model. Several tools have been developed and used for their modeling: Damped Random Walk \citep{kell09, koz17, sansa18}, Gaussian Process (GP) \citep{shap17, kov19, shap19}, Autoencoders \citep{tach20, bank21, sansa21}, Conditional Neural Process \citep{cvor21} and Latent Stochastic Differential Equations (SDEs) \citep{fagi23}. Modeling of quasar light curves could be of great importance for further analysis \citep{jan21} and studying their structure. Deep learning of quasar light curves on a massive scale \citep{kov23} is still improving. Quasar light curves could vary significantly, and their clustering before processing via deep-learning networks, might be beneficial to the performance of the deep-learning quasar light curves process and results \citep{kov23}. Namely, through its training process, SOM groups similar light curves together. Light curves with similar intrinsic variability patterns will be located close to each other on the map, while those that differ significantly (potentially due to observational noise or other factors) will be placed further apart. Grouping light curves of similar patterns together makes the subsequent modeling or analysis with Neural Processes or other methods more effective by focusing on the similar variability patterns of the quasars rather than being misled or obscured by noise. 

In this article, we investigate the potential enhancement of the Conditional Neural Process (QNPy) framework, specifically designed to model quasar light curves \citep{cvor21, pavl24}. We propose an augmentation of this framework by incorporating a preprocessing step that employs clustering through the Self-Organizing Map (SOM) algorithm. The clustering implementation has been executed in Python via the MiniSom library\footnote{More information on Minimalistic implementation of Self Organizing Maps, along with project description and GitHub page could be found on: \url{https://pypi.org/project/MiniSom/}}. We aim to show that SOM-based clustering can improve the QNPy models of the variability of SOM-clustered quasar light curves. 

 The structure of this paper is organized as follows. The ‘Data’ section details the quasar light curves datasets and preprocessing steps. In the ‘Methods’ section, we explain the algorithmic approach and the implementation specifics of the Conditional Neural Process with Self-Organizing Map clustering. The ‘Results’ and 'Conclusion' sections present our findings, interpret the modeling outcomes, and explore their implications. Finally, the ‘Future Work’ section concludes the paper with a recapitulation of the main contributions and a prospective outlook on potential research directions. \\


\section{DATA}

\indent

\begin{figure}
\centerline{\includegraphics[width=1\columnwidth, keepaspectratio]{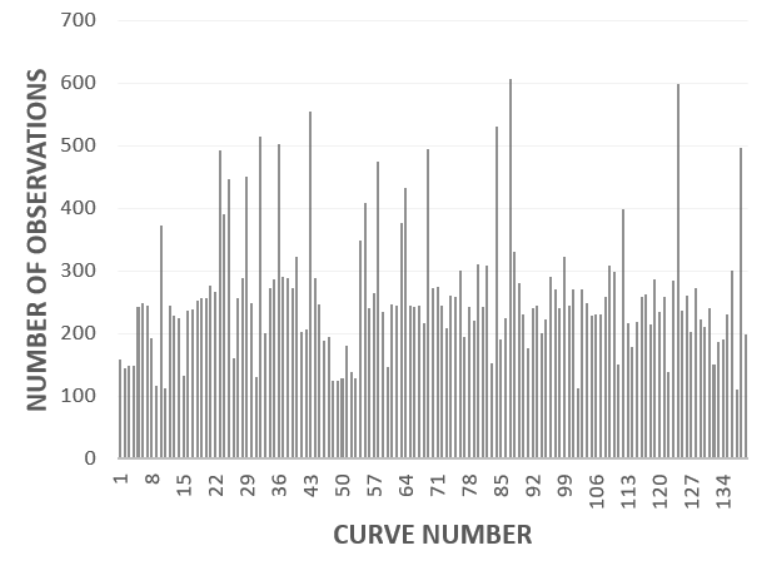}}
\caption{Number of observations in quasar light curves collected in the first nine
months of the all-sky survey by the Swift Burst Alert Telescope.}
\label{fig1}
\end{figure}

The SOM method was trained and tested with quasar light curves, detected by the SWIFT Burst Alert Telescope (9-Month BAT Survey, \citep{tue08}). Their optical light curves were taken from the ASAS-SN (All-Sky Automated Survey for Supernovae) database \citep{hol19}, despite its inherent uncertainties, due to its wide coverage and large dataset. Data points are measured magnitude of selected objects in the g band, in time (modified Julian date). The data set consists of 139 light curves that have between 100 and 600 data points (Figure \ref{fig1}) and a cover range of up to 2000 days.  

Optical light curves taken from the ASAS-SN database \citep{sha14, koch17} have a good representation of 80 percent sky coverage, and show particularities, such as flares, possible quasi-periodic oscillations, time gaps, and irregular cadence, enabling us to test our clustering with challenging data sets. Testing QNPy on light curves with large uncertainties helps assess the robustness and adaptability of the QNPy model. The observational errors have been incorporated as part of the data augmentation strategy to increase the diversity of the training dataset, reducing the risk of overfitting and enhancing the model's ability to generalize to unseen data. Curves from this data set have been used for training and testing the Conditional Neural Process \citep{cvor21}, which was upgraded (QNPy) for modeling the mass amount of curves at once. The need for such a tool comes from an immense amount of data that will be gathered by the next generation time domain surveys, such as Vera Rubin Observatory Legacy Survey of Space and Time (LSST) \citep{abe09, ive19}. This survey will provide observations with different cadences over ten years for millions of AGN sources \citep{bran18, bian21} in six filters - ugrizy. Such a significant amount of data will require efficient tools for processing. Given that QNPy has already been trained and tested on SWIFT/BAT light curves, possible benefits of SOM clustering on the processing efficiency of QNPy, would be most efficiently examined on the same data set.   

The SOM method can handle various data types regardless of their structure and homogeneity, provided each input has a consistent tensor shape (e.g. the same length or number of features). Ensuring all inputs have uniform dimensions, without missing values is crucial for the SOM processing \citep{koh13}. Uniformity of input data is typically achieved through padding or feature extraction \citep{raje22}. These techniques ensure that SOM can process the data, even if the underlying data sources are irregularly sampled or have varying lengths.

Since SWIFT/BAT light curves\footnote{More information on light curve sources can be found on: \url{https://swift.gsfc.nasa.gov/results/bs9mon/}} vary in the number of observations (points), each light curve has been padded by adding the mean magnitude value of that curve at the end, so that all curves have the same number of points as the light curve with the maximum number of points. 
Additionally, curves have been scaled to improve clustering performance. During the testing phase, it was concluded that the best scaler for SWIFT/BAT data was MinMax, which normalized the curves to the interval [0,1]. \\


\section{METHOD}

\indent

\begin{figure}
\centerline{\includegraphics[width=1\columnwidth, keepaspectratio]{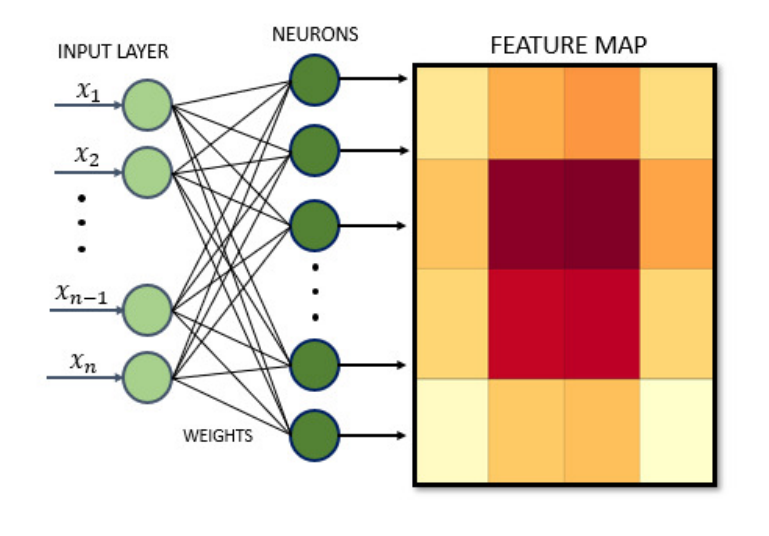}}
\caption{Self-organizing maps are a type of neural network that is trained using unsupervised learning to produce a low-dimensional representation of the input space of the training samples, called a feature map. The feature map in this figure is an output of SOM clustering of SWIFT/BAT data. Values $x_1, x_2,..., x_n$ are vectors (in our case - light curves) that are passed to the input layer.}
\label{fig2}
\end{figure}

Neural networks consist of a collection of connected units or nodes/neurons, which loosely model the neurons in a biological brain. Neurons are grouped to form layers \citep{liu23}. 
Clustering is a type of unsupervised learning, which is used to discover patterns and relationships within data. There are many clustering algorithms: K-Means \citep{ord14}, Gaussian mixtures \citep{tot19}, Spectral clustering \citep{and01}, Hierarchical clustering \citep{yu22}, etc. Among the many machine learning algorithms, Self-Organizing Map (SOM) stands out as a powerful tool for data visualization, clustering, and dimension reduction (Figure \ref{fig2}). The SOM has been used in astronomy mostly for mapping the empirical relation of galaxy color to red-shift \citep{buch19} and visualization and computation of the estimation of galaxy physical properties \citep{hem19}. 

The application of SOMs for clustering quasar light curves presents an advantageous approach due to several characteristics of SOMs. Firstly, SOMs are good at reducing the dimensionality of complex data while preserving topological and metric relationships, making them ideal for handling the high-dimensional nature of quasar light curves. This is particularly beneficial for identifying underlying patterns and structures within the light curves, which may not be readily apparent.
SOMs can capture non-linear and non-stationary variability in quasar light curves by clustering similar light curves based on their intrinsic properties, such as variability amplitude, and time scales of variability. By organizing the light curves into clusters, SOM facilitates a more targeted modeling approach where each cluster represents a different archetype of variability.
Modeling each cluster separately allows for the construction of more specialized and accurate models that can account for the specific characteristics and noise properties of each group. This specificity can lead to more robust predictions and insights into the physical processes governing quasar variability. Moreover, it can improve the efficiency of the modeling process by focusing computational resources on distinct subsets of the data, each with its unique features, rather than applying a one-size-fits-all model to the entire dataset.

The SOM, also known as Kohonen network \citep{koh13} is based on competitive learning. The neurons (or nodes) compete to decide which one will be activated over a set of inputs and this neuron is called the winner or the best matching unit (BMU). 

\begin{figure}
\centerline{\includegraphics[width=1\columnwidth, keepaspectratio]{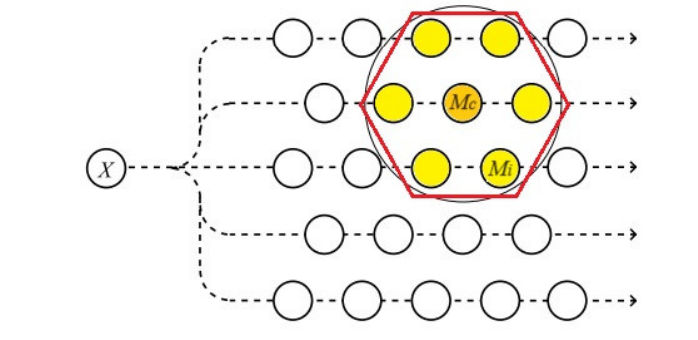}}
\caption{The figure from \citep{koh14} is adapted for our SOM implementation. In the SOM algorithm input data item $X$ is compared to a set of nodes $M_i$. The node $M_c$, which matches the most X is a winning one. Neighboring nodes are more similar to X than nodes outside the circled neighborhood. To determine the winning node, we have used a Gaussian neighborhood function in a rectangular topology.}
\label{fig3}
\end{figure}

Each neuron is connected to a weight vector that has the same dimensions as the input data. At the beginning of the SOM process, the weights are initialized with random values. During forward propagation, input values are compared with nodes (Figure \ref{fig3}) using the "neighborhood function", which calculates the weights of the neighborhood of a position in the map. The Minisom has several options for neighborhood function, but for SWIFT/BAT data Gaussian function (with parameter sigma determining the influence on neighboring nodes of SOM, set to 2.0), showed the best results. During training progress, the algorithm calculates the Euclidean distance between every weight and the current training item: 

\begin{equation}
d(x,w_j) =  \sqrt{\sum_{i=0}^{p-1} (x_i - w_{ij})^2} 
\label{eq:1}
\end{equation}

Vector $x=(x_0,…,x_{p-1})$ presents the input values and $w_{ij}$ weight for
node $j$. Distance determines the difference between two vectors of the same size. The least distance is BMU. The weight vectors are updated by the algorithm to adapt to the distribution of the data by: 

\begin{equation}
w_{ij}(t + 1) = w_{ij}(t) + \theta_j(t)\alpha(t)(x_i(t) - w_{ij}(t)),
\label{eq:2}
\end{equation}

where $i=0,..,p-1$, $t$ shows iteration, $\theta_j(t)$ is the neighbourhood function
and $0<\alpha(t)<1$ that decreases monotonically in time is the learning rate. 
The neighborhood function determines to which extent each output node receives a training adjustment from the current training pattern. The Gaussian function is a common choice for a neighborhood function: 

\begin{equation}
f(x, \mu, \sigma) = \frac{1}{\sigma\sqrt{2\pi}}e^{-\frac{1}{2}(\frac{x - \mu}{\sigma})^2},
\label{eq:3}
\end{equation}

where $x$ represents input, while $\mu$ and $\sigma$ are the mean and the standard deviation (respectively).  

The Minisom method also gives options to choose a topology for clustering, between rectangular and hexagonal, as well as different scalers ('minmax', 'standard', 'robust'), learning rate, etc. After careful testing with various sets of parameters and analyzing the results, it has been concluded that for SWIFT/BAT light curves the best set of parameters is:   

\parindent=.7cm

\par\hang\textindent{(i)}scaler = 'minmax' - normalization function, used in preprocessing phase
\par\hang\textindent{(ii)}neighborhood function = 'gaussian' - neighborhood function, used for updating weights
\par\hang\textindent{(iii)}epochs = 50000 - number of epochs to train SOM
\par\hang\textindent{(iv)}topology = 'rectangular' - topology of SOM
\par\hang\textindent{(v)}learning rate = 0.01
\par\hang\textindent{(vi)}sigma = 2.0 - influence on neighboring nodes of SOM \\

Minmax scaler is developed in sklearn.preprocessing.MinMaxScaler\footnote{More information on MinMaxScaler can be found on: 
\url{https://scikit-learn.org/stable/modules/generated/sklearn.preprocessing.MinMaxScaler.html}} and transforms input to range [0,1]. Transformations are given by the equation: 

\begin{equation}
x_{std} = \frac{x - x_{min}}{x_{max} - x_{min}} 
\label{eq:4}
\end{equation}

\begin{equation}
 x_{scaled} = x_{std}(max - min) + min
\label{eq:5}
\end{equation}

where min and max determine the range in which we want to scale our features (in our case min = 0 and max = 1). 
Clustering of this dataset wasn't computationally demanding. The whole training/clustering process lasted $\approx 5$ minutes.

During testing with various sets of parameters it has been noted that aside from the scaler used for reshaping (preprocessing), topology defining the shape of clusters, and sigma (influence on neighboring nodes of SOM), the SOM is sensitive to outliers in light curves. Small sets of badly clustered curves led to the significantly diminished performance of QNPy. After considering this, additional preprocessing has been done by removing outliers with the Z-Score function: 

\begin{equation}
 z = |\frac{m - \mu}{\sigma}|
\label{eq:6}
\end{equation}

where $m$ is magnitude of observation, $\mu$ is the mean value of magnitudes and $\sigma$ is the standard deviation. Every point (observation) that had $z$ value below $2.0$ was considered an outlier.

The number of clusters is also a hyperparameter in the SOM method \citep{koh13, koh14}. The formula for calculating clusters was fine-tuned through many experiments on different data sets to get an equation suitable for various light curve data. It was important to avoid getting a large number of small clusters, containing very similar curves because that could lead to overfitting during QNPy processing. On the other hand, a very small number of generalized clusters would not improve QNPy performance. 
The length of the input data determines the number of clusters. More specifically, padded curve length specifies the number of grids that create a cluster map. This value is calculated with the following equation: 

\begin{equation}
g = R_u(\sqrt{\sqrt{n}})   
\label{eq:7}
\end{equation}

where $g$ is the number of grids that create a cluster map, $R_u(x)$ is a function that returns the smallest integer greater than or equal to $x$ and $n$ is the length of padded curves.


\section{RESULTS}

\indent

To determine if the clustering of our set of light curves will improve their modeling via the QNPy neural network, the following steps have been taken: 

\parindent=.7cm

\par\hang\textindent{(1)}the whole data set, containing all 139 light curves has been modeled via QNPy, and the results were saved for letter analysis, 
\par\hang\textindent{(2)}the whole data set, containing all 139 curves has been processed via the SOM algorithm. This process included: padding the light curves to maximum length, reshaping them to the interval [0,1] via MinMax scaler, and clustering, 
\par\hang\textindent{(3)}analyzing results and changing parameters, until achieving the best possible results, 
\par\hang\textindent{(4)}selecting several clusters for further analysis, 
\par\hang\textindent{(5)}taking original curves (as they were before padding and reshaping) from each of those clusters and running QNPy only on those curves, 
\par\hang\textindent{(6)}comparing results of modeling for curves before and after clustering.\\

\begin{figure}
\centerline{\includegraphics[width=1\columnwidth, keepaspectratio]{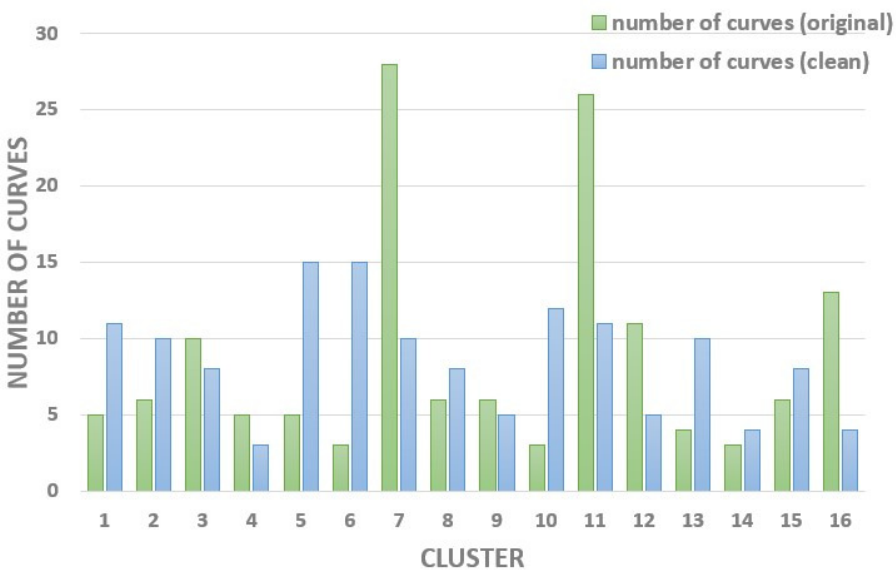}}
\caption{Number of curves in each cluster, before (green) and after (blue) removing outliers from SWIFT/BAT quasar light curves.}
\label{fig4}
\end{figure}

\begin{figure}
\centerline{\includegraphics[width=1\columnwidth, keepaspectratio]{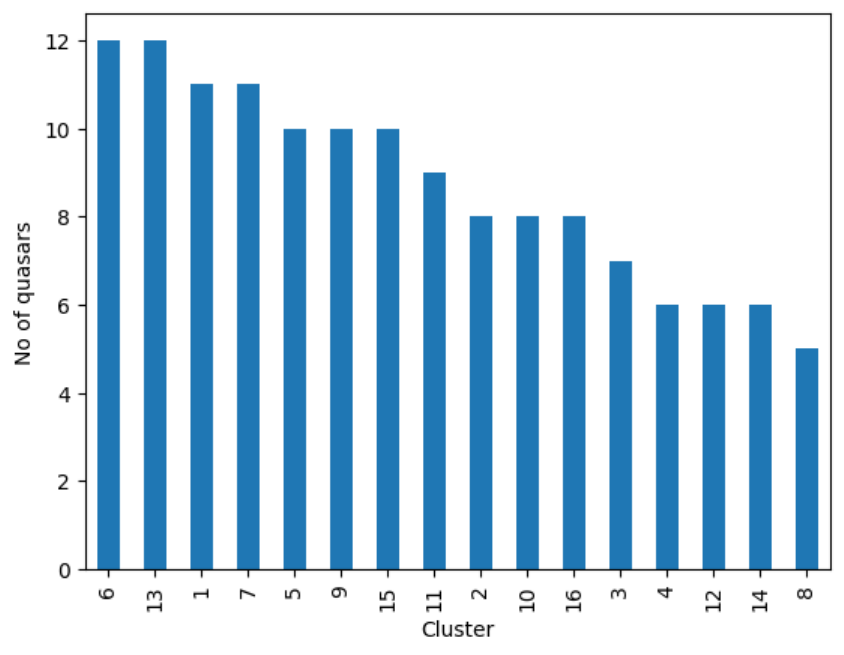}}
\caption{Number of curves in each cluster, after the SOM processing of clean (padded, normalized, removed outliers) SWIFT/BAT quasar light curve data.}
\label{fig5}
\end{figure}

Removing outliers, finding the optimal set of parameters, and careful preprocessing resulted in a more even distribution of light curves in clusters (Figure \ref{fig4}), as well as fewer wrongly clustered curves. The SOM method divided the SWIFT/BAT set into 16 clusters (Figure \ref{fig5}). 

\begin{figure*}[h!]
\centerline{\includegraphics[width=0.85\textwidth, keepaspectratio]{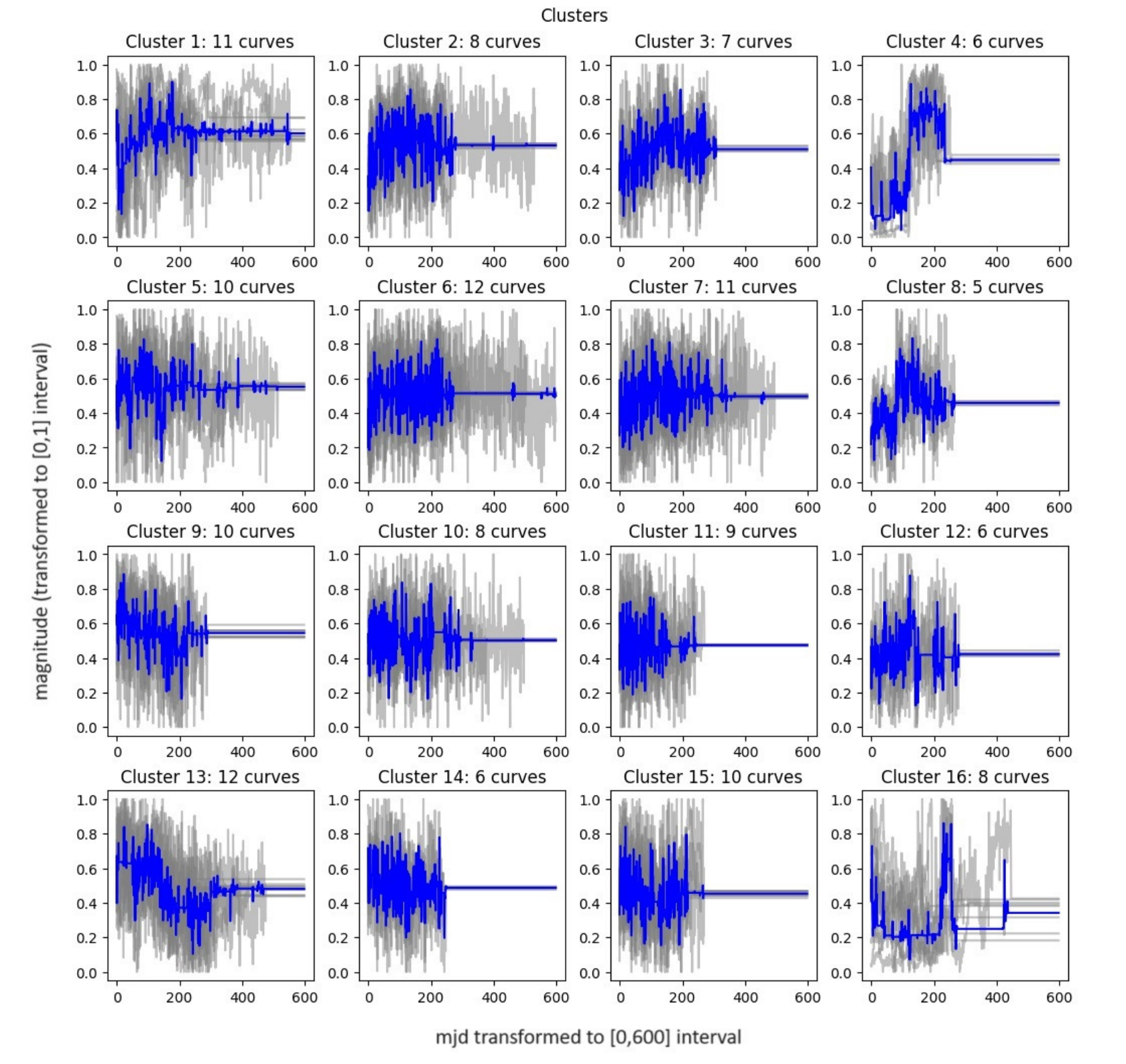}}
\caption{Visualisation of each of 16 clusters, with average curve structure in blue. SOM method requires all curves to be the same length (length of maximum curve). Curves are padded with mean magnitude value. After padding, curves are mapped to [0,1]x[0,600] intervals and clustered.}
\label{fig6}
\end{figure*}

Visualization of curves in clusters (Figure \ref{fig6}) shows the complexity and how the SOM algorithm managed to cluster them according to similarities (gradient changes) in their structure. 
We can see that in some clusters (3, 4, 8, 9, 11, 12, 14) padding is taking a big part of reshaped curves. Although 16 distinct clusters were initially identified  (Figure \ref{fig6}), further analysis demonstrated that these clusters could be aggregated into 4 general groups without significant loss of detail, based on high intra-group similarity and low inter-group similarity: 
\parindent=.7cm

\par\hang\textindent{(1)}Clusters with Low Variability (C-LV): Clusters 4, 10, and 14 exhibit tightly grouped light curves with minimal spread among individual observations, indicative of homogeneous behavior and suggesting lower intrinsic variability or observational noise. 
\par\hang\textindent{(2)}Clusters with Moderate Variability (C-MV): Clusters 2, 5, 6, 9, 12, and 13  demonstrate a moderate level of variability. Despite some spread, individual light curves within these clusters maintain a coherent pattern, categorizing them into an intermediate variability level. 
\par\hang\textindent{(3)}Clusters with High Variability (C-HV): Marked by significant divergence from the mean, Clusters 1, 3, 7, 11, and 15 display pronounced variability. This suggests a high degree of intrinsic quasar variability or substantial observational noise. 
\par\hang\textindent{(4)}Unique Clusters (C-U): Cluster 8 presents a notable trend distinct from other clusters, potentially indicating a different variability type or systematic data effect, thus categorizing it as unique. The 'S' shaped pattern observed in Cluster 8 might indicate quasi-periodic oscillations (QPOs). This oscillatory behavior suggests an underlying periodic process that could be associated with accretion disk dynamics or interactions with the central supermassive black hole's magnetic field, meriting detailed analysis in a separate paper. \\

Moreover, Cluster 16 exhibits a flare-like feature within its mean curve, characterized by a sharp increase in brightness followed by a gradual decrease. This pattern might suggest transient brightening events, which are significant for understanding the dynamics and physical processes within quasars. The presence of such a flare-like appearance warrants further investigation to discern the nature of these transient events and their astrophysical implications. 
Additionally, some clusters consist of curves that were padded with very different numbers of points (2, 5, 6). This could have influenced the clustering of padded curves. To determine if clustering would improve the QNPy modeling of the SWIFT/BAT data set, the following clusters were selected based on several reasons (Table \ref{table1}), which should cover different types and give an overview of the pros and cons of the SWIFT/BAT data set SOM clustering.  

\begin{table*}
\caption{Clusters selected for analysis of SOM clustering effect on QNPy modeling performance.}
\parbox{\textwidth}{
\vskip.25cm
\centerline{\begin{tabular}{@{\extracolsep{0.0mm}}lcccccccccccc@{}} 
\hline
    Cluster number & Number of curves & Selection reason  \\
\hline
 4 (C-LV type )& 6 &a small number of curves and a peculiar structure with one big gradient change\\
 13 (C-MV type) & 12 & the largest number of variable curves, the mean model has a visible structure\\
 1 (C-HV type) & 11 & highly variable curves, with different degrees of padding\\
 8 (C-U type) & 11 &  distinct from other clusters potentially indicating different variability types\\
 16 (specific) & 8 & the mean model has a very peculiar flare-like feature within its mean curve\\  
\hline
\end{tabular}}}
\label{table1}
\end{table*}

\begin{figure*}[h!]
\centerline{\includegraphics[width=0.90\textwidth, keepaspectratio]{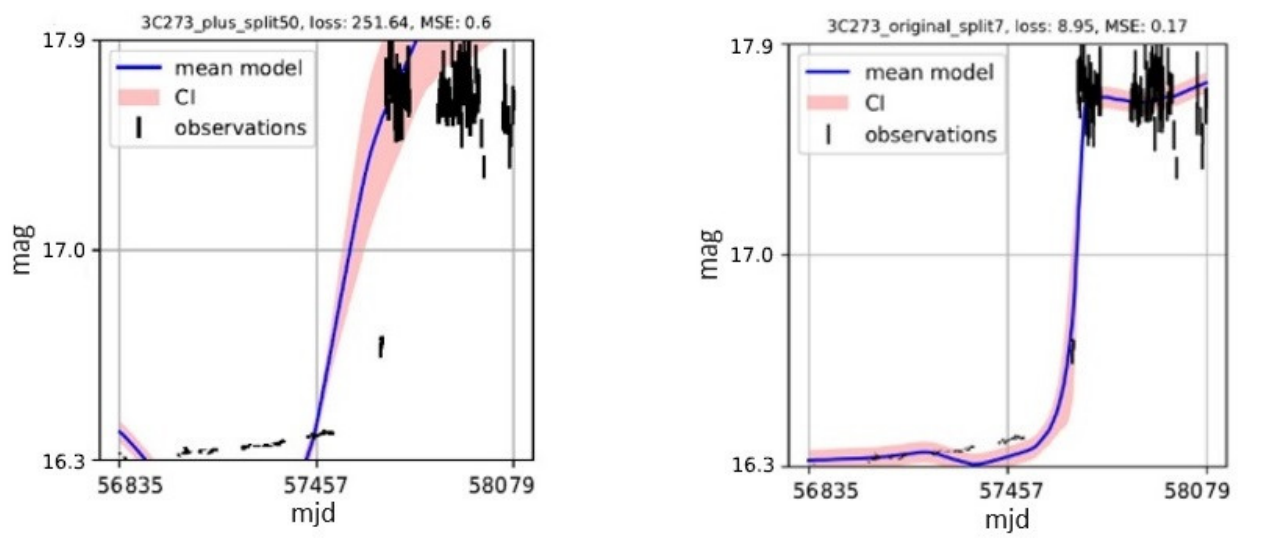}}
\caption{The QNPy modeling of 3C273 light curve using 3000 iterations. Data is transformed to [-2,2][-2,2] interval during the QNPy preprocessing phase. The light red area is the one-sigma confidence interval. The left graph shows the results of the modeling of 3C273 during the processing of the whole data set, while the right side graph shows the modeling of 3C273 during the processing of Cluster 4. Cluster 4 is an example of a cluster with low variability (C-LV). The mean model is shown as a blue line.}
\label{fig7}
\end{figure*}

Cluster 4 is an example of a cluster with Low Variability (C-LV). This cluster consists of 6 curves with very similar structures and distinctive slopes. Figure \ref{fig7} shows the difference in modeling when 3C273 is modeled as part of the whole data set (left) and as a part of Cluster 4 (right). Loss function and mean squared error (MSE) are much smaller when QNPy processes only Cluster 4. Both modeled curves (blue lines) follow the main gradient change, but after clustering the model goes nicely through the mean of data points on the lower and upper part of the curve. 

\begin{figure*}[h!]
\centerline{\includegraphics[width=0.90\textwidth, keepaspectratio]{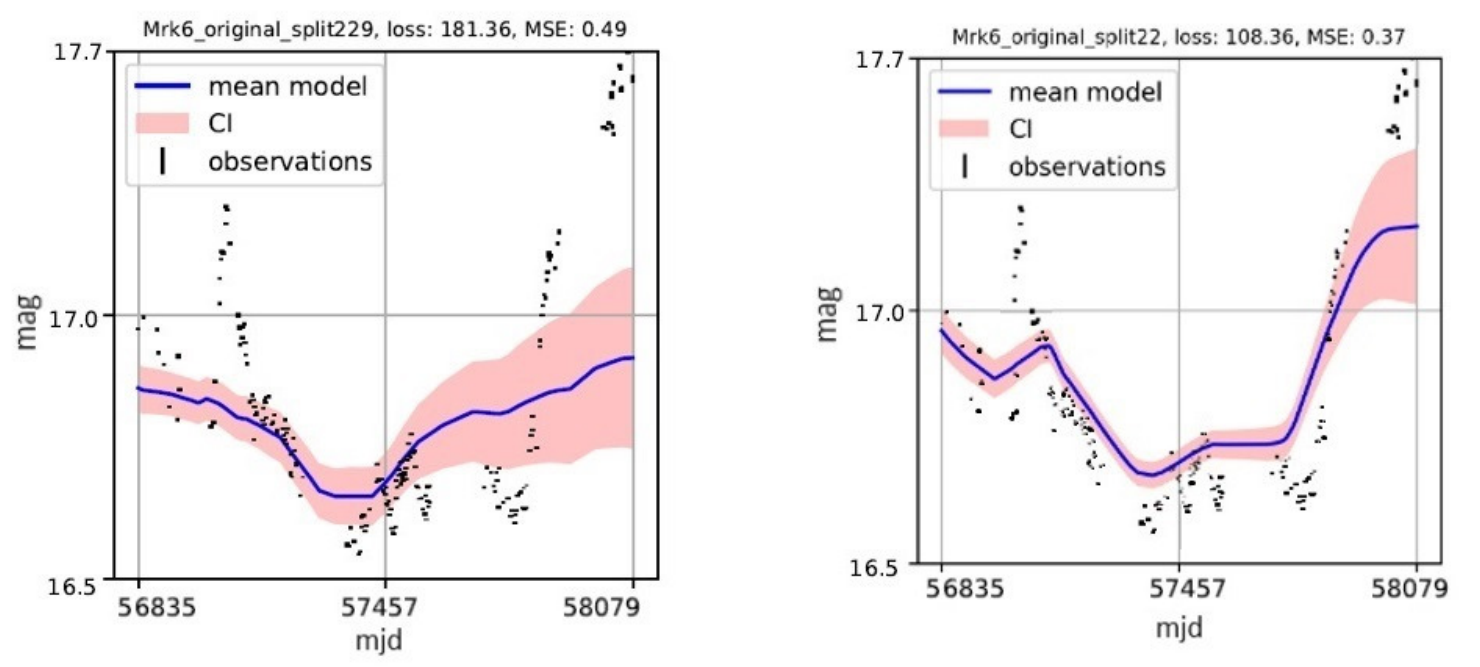}}
\caption{the QNPy modeling of Mrk6 light curve using 3000 iterations. Data is transformed to [-2,2][-2,2] interval during the QNPy preprocessing phase. The light red area is the one-sigma confidence interval. The left graph shows the results of the modeling of Mrk6 during the processing of the whole data set, while the right side graph shows the modeling of Mrk6 during the processing of Cluster 13. Cluster 13 is an example of a cluster with moderate variability (C-MV). The mean model is shown as a blue line.}
\label{fig8}
\end{figure*}

Cluster 13 is an example of a cluster with Moderate Variability (C-MV). This cluster consists of 12 curves with very similar structures and distinctive gradient changes. Figure \ref{fig8} shows the difference in modeling when Mrk6 is modeled as part of the whole data set (left) and as a part of Cluster 13 (right). Loss function and mean squared error (MSE) are smaller when QNPy processes only Cluster 16. Both modeled curves (blue lines) follow gradient changes, but after clustering the model fits data better through the whole length. While the first modeled curve manages to model gradient change in the middle, it stays near the mean. The second curve nicely follows the whole pattern of the curve, but still does not reach the top of the first maximum and the minimal values in the middle.  

\begin{figure*}[h!]
\centerline{\includegraphics[width=0.90\textwidth, keepaspectratio]{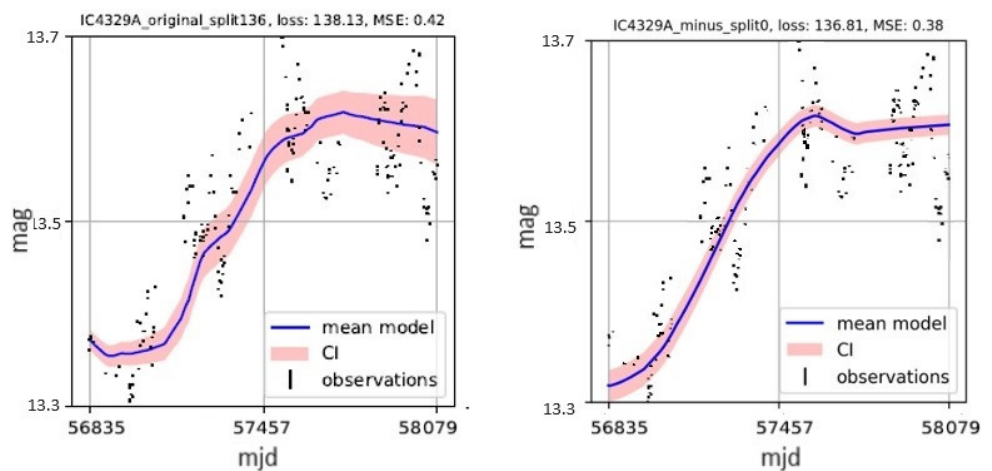}}
\caption{the QNPy modeling of IC4329A light curve using 3000 iterations. Data is transformed to [-2,2][-2,2] interval during the QNPy preprocessing phase. The light red area is the one-sigma confidence interval. The left graph shows the results of the modeling of IC4329A during the processing of the whole data set, while the right side graph shows the modeling of IC4329A during the processing of Cluster 1. Cluster 1 is an example of a cluster with moderate variability (C-HV). The mean model is shown as a blue line.}
\label{fig9}
\end{figure*}

Cluster 1 is an example of a cluster with High Variability (C-HV). This cluster consists of 11 curves with very variable structures. Figure \ref{fig9} shows the difference in modeling when IC4329A is modeled as part of the whole data set (left) and as a part of Cluster 1 (right). Loss function and mean squared error (MSE) are larger when QNPy processes only Cluster 1. Both modeled curves (blue lines) follow the mean of the main gradient change, but neither of them captures details of the light curve. While the first modeled curve shows small gradient changes in more dense regions, the second one is completely smooth. This indicates that for clusters that contain highly variable curves, modeling results are close to the modeling of the whole set, which is still impressive given the significantly smaller number of curves for learning, making it much more efficient (considering the time it takes to model this cluster is several minutes, whereas for the whole set on the same machine takes a couple of hours).  

\begin{figure*}[h!]
\centerline{\includegraphics[width=0.90\textwidth, keepaspectratio]{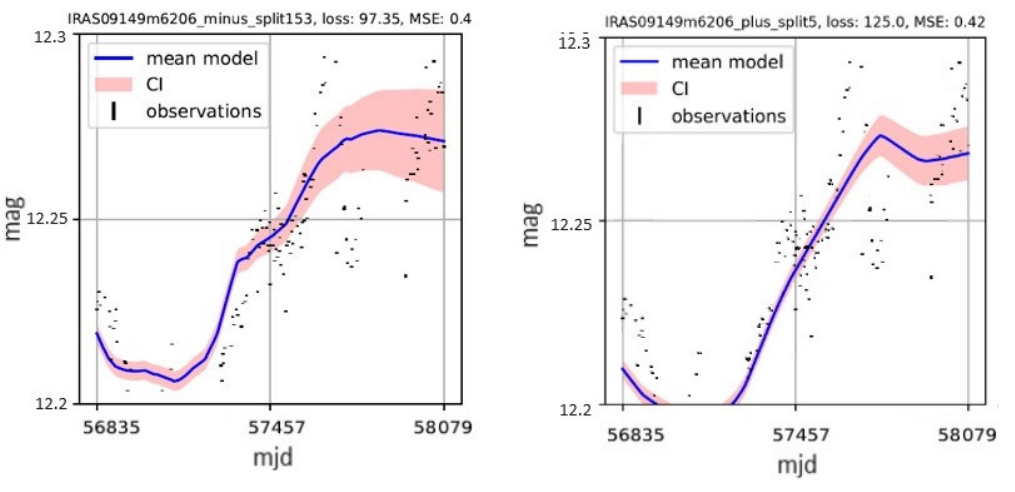}}
\caption{the QNPy modeling of IRAS09149m6206 light curve using 3000 iterations. Data is transformed to [-2,2][-2,2] interval during the QNPy preprocessing phase. The light red area is the one-sigma confidence interval. The left graph shows the results of the modeling of IRAS09149m6206 during the processing of the whole data set, while the right side graph shows the modeling of IRAS09149m6206 during the processing of Cluster 8. Cluster 8 is an example of a Unique Cluster (C-U). The mean model is shown as a blue line.}
\label{fig10}
\end{figure*}

Cluster 8 is an example of a Unique Cluster (C-U). This cluster consists of 5 curves with very distinctive structures. Figure \ref{fig10} shows the difference in modeling when IRAS09149m6206 is modeled as part of the whole data set (left) and as a part of Cluster 8 (right). Loss function and mean squared error (MSE) are higher when QNPy processes only Cluster 8. Both modeled curves follow the main gradient of the light curve, but the first case (modeling of the whole set) has more details and small gradient changes in the more dense regions. The second graph (modeling curves from cluster 8) is a smooth representation of the main data trend. This indicates that for clusters that contain unique curves (some possibly wrongly clustered), especially if they consist of very few examples (6 curves), modeling results are less accurate than those of modeling the whole set.  

\begin{figure*}[h!]
\centerline{\includegraphics[width=0.90\textwidth, keepaspectratio]{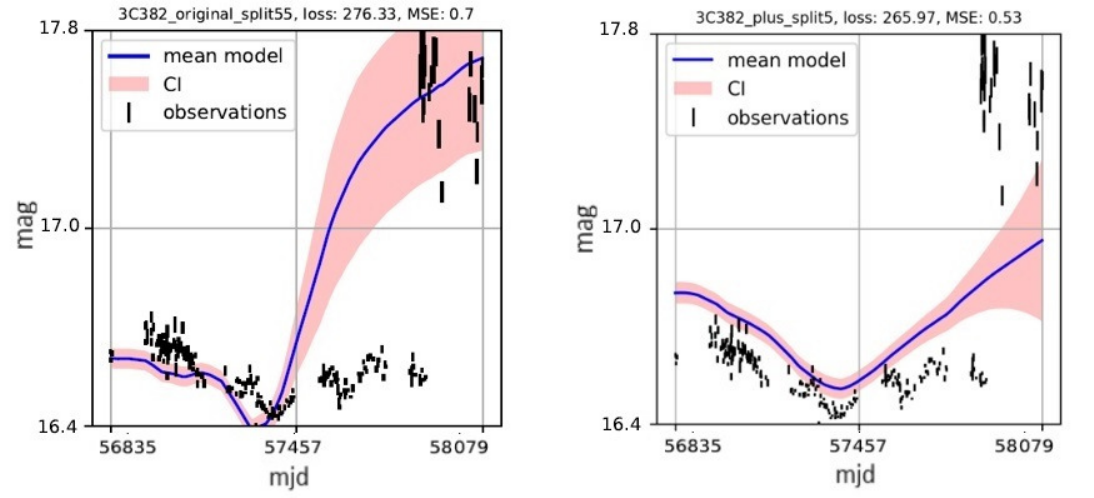}}
\caption{the QNPy modeling of 3C382 light curve using 3000 iterations. Data is transformed to [-2,2][-2,2] interval during the QNPy preprocessing phase. The light red area is the one-sigma confidence interval. The left graph shows the results of the modeling of 3C382 during the processing of the whole data set, while the right side graph shows the modeling of 3C382 during the processing of Cluster 16. The mean model is shown as a blue line.}
\label{fig11}
\end{figure*}

Cluster 16 consists of 8 curves and a very peculiar mean model. After further analysis of light curves that were clustered in this cluster, it was apparent that some of them were wrongly clustered. The most obvious example was a light curve 3C382 (Figure \ref{fig11}), which should have been clustered in Cluster 4. After running QNPy on Cluster 16, the results varied from model to model. Figure \ref{fig11} shows that modeling of the whole data set gives better results if clustering is not accurate. 

The first model follows the main gradient change and overall data structure while the second one stays closer to the mean value following gradient changes of the first part of the curve. It is also visible that loss and MSE could have lower values for a less accurate model. This could be caused by the fact that testing and validation sets are very small for clustered curves and there is a larger probability that they will randomly get the curves that fit the model better. This indicates that there is still a need for further improvement of the clustering algorithm. \\

\section{CONCLUSION}

\indent

Quasars are one of the most powerful sources in the Universe. Their radiation covers the whole electromagnetic spectrum in a very complex manner. The gathering and analysis of quasar light curves have been revealing the structure of these complex objects and continue to do so with the development of more powerful observing technologies. We need to perfect our programming tools to keep up with the immense amount of data that the most powerful telescopes can now harbor. Conditional Neural Process, which has shown to be a good modeling tool for capturing variability and complexity of quasar light curves, still struggles with modeling a large number of these complex data sets. To improve its performance, we examined SOM clustering as a means of preprocessing and selecting batches of quasar light curves, which would give better and faster results.  

SWIFT/BAT data was selected since it gives a good starting point for determining the applicability of SOM clustering on complex, stochastic, irregular, and very different sources. This data set consists of 139 very complex, variable, inhomogeneous light curves with many gaps and different structures and lengths. The same data set was used for training and testing the Conditional Neural Process on which QNPy has been built. Testing Neural Processes on light curves with large uncertainties helps assess the robustness and adaptability of the QNPy model. 

The SOM algorithm was selected for this task since it is not a very complex machine-learning algorithm for unsupervised learning and it works fast, which is essential for preprocessing purposes. It is trained through a competitive neural network (neurons in the output layer compete among themselves to be activated). The competitive process is to search for the most similar neuron (BMU) with the input pattern. The BMU and its neighbors adjust their weights. We have used MiniSom implementation of the Self Organizing Maps (SOM) and attempted to make the best clustering of SWIFT/BAT curves. Curves from several selected clusters were modeled via QNPy and results were compared with the original clustering of the whole set.

It was noted that the SOM parameters (neighborhood function, topology, scaler, etc) contribute significantly to the clustering process and results. Additionally, the SOM method required the reshaping of light curves via normalization (minmax) and transformation to the same interval. Removing outliers in light curves is beneficial for better clustering and a more even distribution of light curves among clusters. 

The results indicate that curves from clusters whose mean value has distinctive features, with low and medium variability (C-LV, C-MV), show more improvement in modeling via QNPy than those with highly variable mean functions (C-HV) or unique structures(C-U), especially if such cluster contains a small number of curves. Comparing the results of modeled light curves before and after clustering, showed that good clustering could contribute to greater extant performance of the QNPy modeling process. However, the SOM has shown to be very sensitive to some data features which could lead to wrongly clustered curves. In the case of wrongly clustered light curves, the QNPy modeling shows far less reliable results than the modeling of the full set of data. Not only because there is less data to "learn" from, but because most data has certain features that are not present in wrongly clustered curves light curves. 

The clustering of quasar light curves as a means of preprocessing could significantly improve modeling via QNPy, but clustering itself could be improved to achieve the most optimal results. The SOM could be used for this task, but it needs further analysis and testing on various datasets, with careful tuning of its parameters and good preparation of input data.  \\


\section{FUTURE WORK}

\indent

The SOM algorithm could be tested on a set of light curves that have a similar number of data points in the same period so that reshaping and padding don't change the structure of light curves before clustering. 
ZTF (Zwicky Transient Facility) data would be a good choice for this analysis since it has fewer uncertainties, which makes it a good sample for further exploring the effectiveness of QNPy and creating a comprehensive analysis, offering deeper insights into light curve behaviors.

Padding could be done via interpolation, reflective padding, or replication. The selection of the number of clusters could be explored further. Clustering could be done via other methods to determine the one that would make the best clustering results and hence contribute more to the improvement of the QNPy process.


\acknowledgements{This paper is greatly improved by the contribution, guidance, and consultation from Dr. A. Kova{\v c}evi{\' c} and Dr. M. Pavlovi{\' c}, along with the highly appreciated comments from referees. The author acknowledges the funding provided by the University of Belgrade - Faculty of Mathematics (the contract 451-03-66/2024-03/200104) through the grants by the Ministry of Science, Technological Development and Innovation of the Republic of Serbia.}


\vskip2mm

\newcommand\eprint{in press }

\bibsep=0pt

\bibliographystyle{aa_url_saj}

{\small

\bibliography{sample_saj}
}


\begin{strip}

\end{strip}

\clearpage

{\ }

\clearpage

{\ }

\newpage

\begin{strip}

{\ }



\naslov{MODELOVANjE PROMENLJIVOSTI KVAZARA UPOTREBOM NEURONSKIH PROCESA BAZIRANIH NA SAMOORGANIZUJU{\CC}IM MAPAMA}


\authors{I. {\v C}vorovi{\' c} - Hajdinjak$^{1}$}

\vskip3mm


\address{$^1$Department of Astronomy, Faculty of Mathematics,
University of Belgrade\break Studentski trg 16, 11000 Belgrade,
Serbia}


\Email{iva.cvorovic@gmail.com}

\vskip3mm


\centerline{{\rrm UDK} \udc}


\vskip1mm

\centerline{\rit Uredjivaqki prilog}

\vskip.7cm

\baselineskip=3.8truemm

\begin{multicols}{2}

{
\rrm

Kondicionalni neuronski procesi su se pokazali kao mo{\cc}ni alati za modelovanje svetlosnih krivih kvazara. Imaju{\cc}i u vidu kompleksnu prirodu ovih objekata, a samim tim i podataka koji nam od njih sti{\zz}u, modelovanje svetlosnih krivih mo{\zz}e biti veoma vremenski zahtevno. Rezultati modelovanja nekada nisu dovoljno precizni da bi se na njima vr{\ss}ila dalja analiza. U cilju unapredjenja modelovanja svetlosnih krivih, u ovom radu je ispitan efekat klasterovanja velikog broja svetlosnih krivih kvazara kori{\ss}{\cc}enjem samoorganizuju{\cc}ih mapa (SOM), u procesu koji bi prethodio neuronskom modelovanju. Rezultati primene SOM metode na skupu svetlosnih krivih kvazara, prikupljenih u programu ASAS-SN, i njihovo kasnije modelovanje, pokazju da kondicionalni neuronski procesi rade bolje nakon prethodnog procesiranja podataka SOM metodom. Zaklju{\ch}ak ovog rada je da klasterovanje krivih kvazara SOM metodom, pre modelovanja kondicionalnim neuronskim procesima, unapredjuje rezultate modelovanja.

{\ }

}

\end{multicols}

\end{strip}



\end{document}